\documentclass[conference]{IEEEtran}
\usepackage{amsfonts}
\usepackage{times}
\usepackage{graphicx}
\usepackage{latexsym}
\usepackage{dsfont}
\usepackage{amssymb}
\usepackage{amsmath}
\usepackage{cite}
\usepackage{verbatim}
\usepackage{subfigure}

\newcommand{\figref}[1]{{Fig.}~\ref{#1}}

\def\bb0{{\mathbb{0}}}

\def\bb{{\mathbf{b}}}

\def\bff{{\mathbf{f}}}

\def\b0{{\mathbf{0}}}

\def\bX{{\mathbf{X}}}

\def\sf0{{\mathsf{0}}}

\usepackage{epstopdf}
\usepackage{enumerate}
\usepackage{algorithmicx}
\usepackage{algorithm}
\usepackage{amsmath}
\usepackage[noend]{algpseudocode}
\usepackage{float}
\usepackage{hyperref}
\usepackage{color}
\usepackage{makeidx}
\usepackage{bbm}
\usepackage{graphicx}
\usepackage{lipsum}
\usepackage{subfigure}
\usepackage{tablefootnote}
\usepackage{multirow}
\usepackage{multicol}
\usepackage{balance}
\usepackage{booktabs}
\DeclareGraphicsExtensions{.eps, .pdf}

\newcommand{\sref}[1]{{Section}~\ref{#1}}

\newcommand{\comm}[1]{}

\begin{document}

\title{Millimeter Wave Drones with Cameras: \\ Computer Vision Aided  Wireless Beam Prediction }
\author{Gouranga Charan, Andrew Hredzak, and Ahmed Alkhateeb\\ Arizona State University, Emails: \{gcharan, ahredzak, alkhateeb\}@asu.edu}

\maketitle

\begin{abstract}
Millimeter wave (mmWave) and terahertz (THz) drones have the potential to enable several futuristic applications such as coverage extension, enhanced security monitoring, and disaster management. However, these drones need to deploy large antenna arrays and use narrow directive beams to maintain a sufficient link budget. The large beam training overhead associated with these arrays makes adjusting these narrow beams challenging for highly-mobile drones. To address these challenges, this paper proposes a vision-aided machine learning-based approach that leverages visual data collected from cameras installed on the drones to enable fast and accurate beam prediction. Further, to facilitate the evaluation of the proposed solution, we build a synthetic drone communication dataset consisting of co-existing wireless and visual data. The proposed vision-aided solution achieves a top-$1$ beam prediction accuracy of $\approx 91\%$ and close to $100\%$ top-$3$ accuracy. These results highlight the efficacy of the proposed solution towards enabling highly mobile mmWave/THz drone communication. 
\end{abstract}


\section{Introduction} \label{sec:Intro}
Drones (and unmanned aerial vehicles (UAVs)) \cite{bariah2020drone} are envisioned to play a critical role in enabling futuristic applications such as extending the coverage of mmWave/THz wireless networks, supporting latency-critical applications, and enabling security monitoring systems. To satisfy the high data rate requirements of these novel application, the drones are expected to be equipped with mmWave/THz transceivers \cite{Rappaport2019}. This is primarily due to the large bandwidth offered by the mmWave/THz communication systems. However, these systems need to deploy large antenna arrays and use narrow directive beams to guarantee sufficient receive SNR. Selecting the optimal beams in these high-frequency systems with large antenna arrays is typically associated with large beam training overhead. This high beam training overhead makes it difficult for these systems to compute accurate beams frequently, making it challenging to support highly-mobile drones. This motivates looking for new approaches to overcome the challenges and enable highly-mobile mmWave/THz drone communication.

Developing solutions to overcome the beam training overhead challenge in mmWave/THz systems has gathered considerable interest in recent years \cite{Hur2011,Alkhateeb2014, RobertPos, Joao2022a, Charan2022a, charan2022c, Jiang2022a, demirhan2022beam, radar2022mag}. Initial approaches focused on the following: (i) Beam training with adaptive beam codebook \cite{Hur2011} and (ii) compressive channel estimation by leveraging channel sparsity \cite{Alkhateeb2014}. An exhaustive or adaptive beam training was proposed in \cite{Hur2011} to obtain the optimal beam at the transmitter and receiver. In \cite{Alkhateeb2014}, the authors propose to leverage the inherent sparsity in mmWave channels and formulate the mmWave channel estimation as a sparse reconstruction problem. Although these classical approaches can help reduce the beam training overhead, they can typically save only one order of magnitude in the training overhead, which is not sufficient to support highly-mobile multi-user scenarios. This, further, motivated the development of machine learning-based solutions that can leverage prior observations and additional sensing data such as user position \cite{RobertPos, Joao2022a}, camera/visual images \cite{Charan2022a, charan2022c}, LiDAR point-cloud data \cite{Jiang2022a}, and radar data \cite{demirhan2022beam, radar2022mag} to name a few. However, the solutions are based on scenarios with humans, vehicles, or robots acting as the user equipment (UE), where the UE's motion is generally restricted to two dimensions and is relatively easy to predict.  

The movement of UAVs, specifically the high mobility along with the multiple possible orientations, imposes a unique challenge for accurate beam prediction. This has further motivated several studies on this specific problem of beam management in drones/UAVs \cite{Song2021a, Yuan2020a }. These solutions propose to utilize user-side (drone) information, such as the position, angle between the basestation and drone, etc., to develop solutions to overcome these challenges and accurately predict current and future beams. The position-aided solutions, although promising, might not scale to real-world scenarios with inherent non-idealities. For example, relying only on location alone might result in inaccurate predictions due to the inherent errors associated with the position (GPS) data.

\begin{figure*}[!t]
	\centering
	\includegraphics[width=\linewidth]{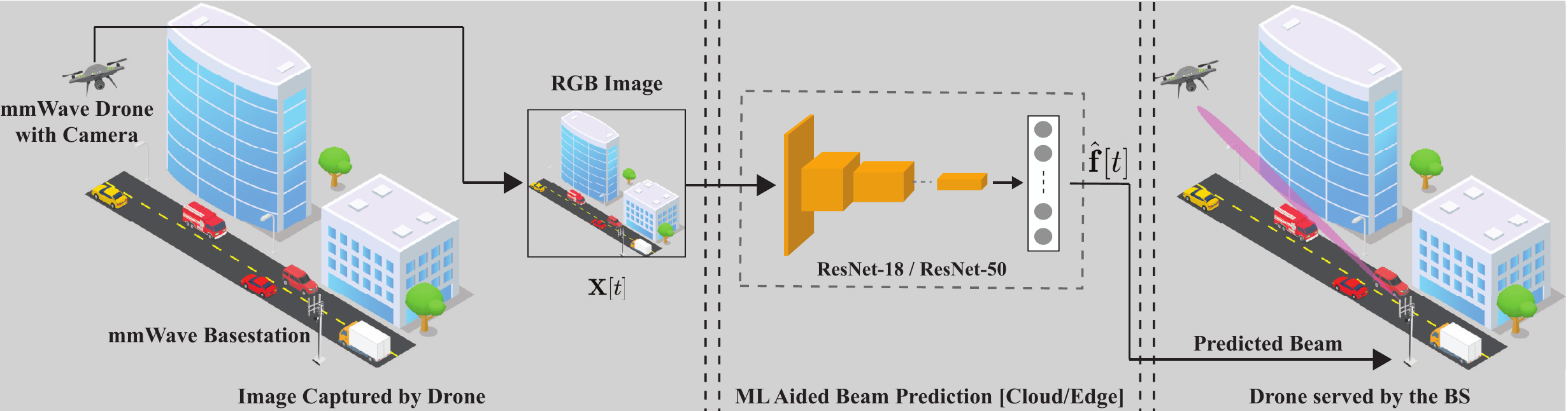}
	\caption{A block diagram showing the proposed solution for the vision-aided beam prediction task. As shown in the figure, the mmWave drone equipped with a camera captures real-time images of the wireless environment. A CNN is then utilized to predict the optimal beam index.  }
	\label{fig:beam_pred_soln}
\end{figure*}

This paper proposes a deep learning-based vision-aided solution to reduce the beam training overhead in mmWave/THz drone communication. The main contribution of this work can be summarized as follows:
\begin{itemize}
	\item {Formulating the vision-aided beam prediction problem for mmWave/THz drone communication considering practical visual and communication models.}
	\item{Developing a deep learning based solution for mmWave/THz drone beam prediction using visual data captured by cameras installed on the drones.  }
	\item{Developing a new dataset for the vision-aided drone beam prediction based on the publicly available ViWi \cite{ViWi} framework. The ViWi-Drone dataset consists of co-existing multi-modal visual and wireless data.}
\end{itemize}
The proposed vision-aided solution achieves more than $90\%$ top-1 beam prediction accuracy on the ViWi-Drone dataset. This highlights the potential of using additional sensory data such as visual images in reducing the beam training overhead.


\section{System Models} \label{sec:sys_model}

This work considers a communication system deployed in a realistic downtown location where a mmWave basestation is serving a flying drone. This section presents the system model adopted in this work. This paper adopts the system model where a basestation equipped with an $M$-element uniform linear array (ULA) is serving a mmWave drone. The drone carries a single-antenna mmWave receiver and is equipped with three RGB cameras to capture the wireless environment. The communication system adopts OFDM transmission with K subcarriers and a cyclic prefix of length D. The basestation is assumed to employ a pre-defined beamforming codebook $\boldsymbol{\mathcal F}=\{\mathbf f_q\}_{q=1}^{Q}$, where $\mathbf{f}_q \in \mathbb C^{M\times 1}$ and $Q$ is the total number of beamforming vectors. Let $\mathbf h_{k}[t] \in \mathbb C^{M\times 1}$ denote the downlink channel between the basestation and the drone at the $k$th subcarrier and time $t$. The received signal at the drone can then be expressed as 
\begin{equation}\label{eq:sys_mod}
	y_{k}[t] = \mathbf h_{k}^T[t] \mathbf f_q[t]x + v_k[t],
\end{equation}
where $\mathbf f \in \boldsymbol{\mathcal F}$ is the optimal beamforming vector at time $t$ and $v_k[t]$ is a noise sample drawn from a complex Gaussian distribution $\mathcal N_\mathbb C(0,\sigma^2)$. The transmitted complex symbol $x\in \mathbb C$ needs to satisfy the following constraint $\mathbb E\left[ |x|^2 \right] = P$, where $P$ is the average symbol power. The beamforming vector $\mathbf f^{\star}[t] \in \boldsymbol{\mathcal F}$ at each time step $t$ is selected to maximize the average receive SNR and is defined as 
\begin{equation}\label{eq:beam_training}
	\mathbf f^{\star}[t] = \underset{\mathbf f_q[t]\in \mathcal F}{\text{argmax}} \frac{1}{K}\sum_{k=1}^{K} \mathsf{SNR}|\mathbf h_{k}^T[t] \mathbf f_q[t] |^2,
\end{equation}
where $\mathsf{SNR}$ is the transmit signal-to-noise ratio, SNR = $\frac{P}{\sigma^2}$.

\begin{figure*}[!t]
	\centering
	\includegraphics[width=\linewidth]{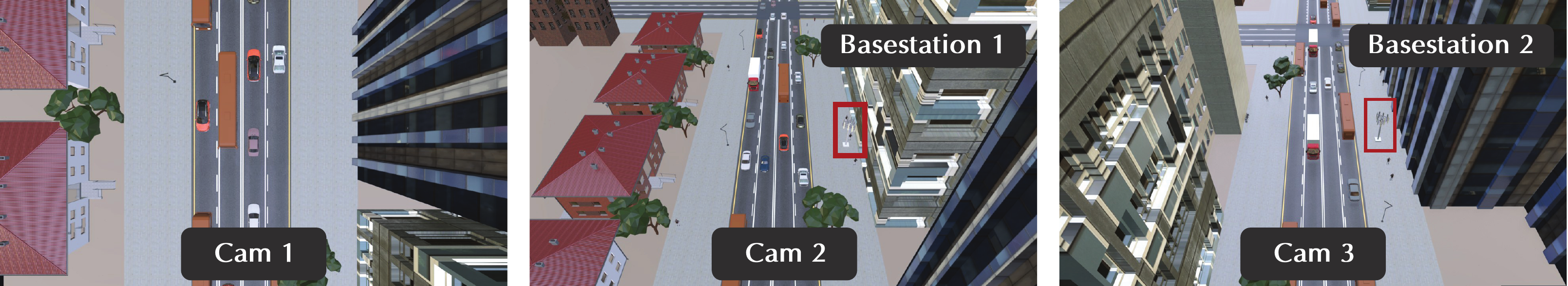}
	\caption{The figure shows the different regions of the street captured by the three cameras at a particular time instance. It also shows the two basestations, BS1 and BS2 as seen from cameras 2 and 3, respectively.   }
	\label{fig:img_samples}
\end{figure*}
\begin{figure}[!t]
	\centering
	\includegraphics[width=\linewidth]{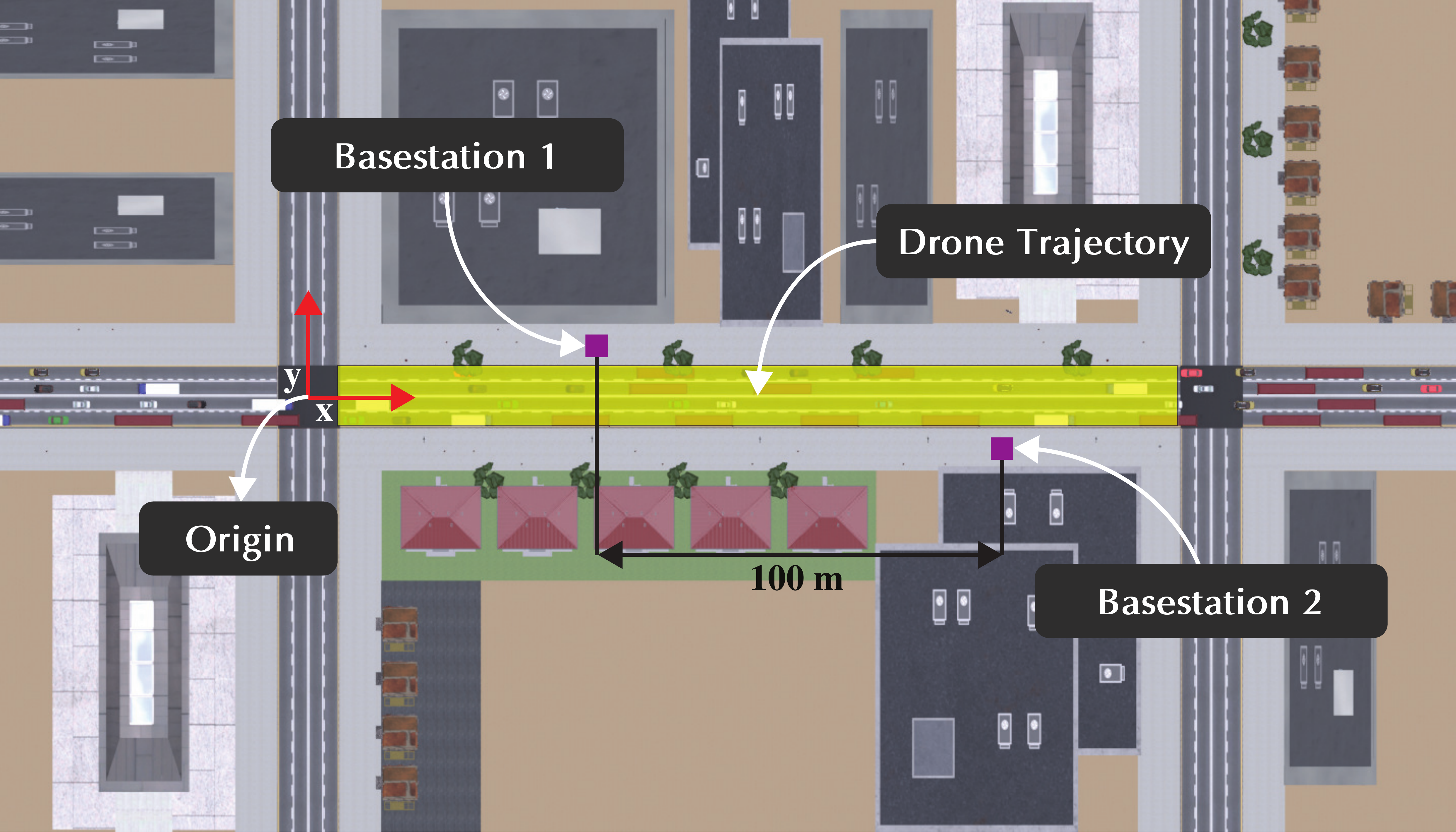}
	\caption{A top-view of the ViWi-Drone scenario. It is modeled after a busy downtown street with diverse objects, such as cars, buses, trucks, etc. It also highlights the trajectory of the drone adopted in this dataset. }
	\label{fig:drone_trajectory}
\end{figure}

\section{Problem formulation} \label{sec:prob_form}
Given the system model in Section~\ref{sec:sys_model}, the beam prediction task can be defined as selecting the best beamforming vector $\bff^\star$ from the codebook $\boldsymbol{\mathcal F}$ such that the receive SNR is maximized. Computing the optimal beams in mmWave/THz communication systems require explicit channel knowledge $\mathbf h_{k}$ (as presented in \eqref{eq:beam_training}. However, it is generally hard to acquire the channel information for these systems. The other alternative is to perform exhaustive search over the beam codebooks. But, given that the mmWave/THz systems need to adopt large antenna arrays and use narrow directive beams, the exhaustive search approach incurs large beam training overhead. All these makes it challenging for the mmWave/THz systems to support highly-mobile drones. Drones are typically equipped with an suite of sensors such as high-resolution camera, IMUs, GPS receivers, etc. In this paper, we propose to predict the optimal beam index by utilizing the visual data collected by the drone. Formally, we define $\bX[t] \in \mathbb{R}^{W \times H \times C}$ as the corresponding RGB image, captured by the camera installed in the drone at time $t$, where $W$, $H$, and $C$ are the width, height, and the number of color channels of the image. The objective of this beam prediction task is to find a prediction/mapping function $f_{\Theta}$ that utilizes the visual data captured by the mmWave drones to predict (estimate) the optimal beam index $ \hat{\mathbf f}[t] \in \boldsymbol{\mathcal F}$. The mapping function can be formally expressed as
\begin{equation}
	f_{\Theta}: \bX[t] \rightarrow  \hat{\mathbf f}[t].
\end{equation}
In this work, we develop a machine learning model to learn this prediction function $f_{\Theta}$. Let $ \mathcal D = \left\lbrace \left (\bX_u, \mathbf f^{\ast}_u \right) \right\rbrace_{u=1}^U $ represent the available dataset consisting of image-beam pairs, where $U$ is the total number of samples in the dataset. The set $\Theta$ in the prediction function represents the model parameters and is learned from the dataset $ \mathcal D$ of labeled data samples. Then, the objective is to maximize the number of correct predictions over all the sample in $ \mathcal D$. This can be formally written as 
\begin{equation}\label{eq:prob_form_1}
	f^{\star}_{\Theta^{\star}} = \underset{f_{\Theta}}{\text{argmax}}\\ \prod_{u=1}^U \mathbb P\left( \hat{\mathbf f}_u = \mathbf f^{\star}_u | \bX_u \right),
\end{equation}
where the product in \eqref{eq:prob_form_1} is due to the implicit assumption that the samples in the dataset $ \mathcal D$ are drawn from an independent and identically distribution (i.i.d.). The next section presents the proposed machine learning model for vision-aided mmWave/THz drone beam prediction.

\section{Proposed Camera Aided Solution} \label{sec:prop_sol}

This work proposes to utilize the visual data captured by the mmWave drones to predict the downlink (from the basestation to the drone) optimal beam indices. In this section, we first explain the key idea in Section~\ref{subsec:key_idea} of the proposed vision-aided solution and then present an in-depth overview of the proposed solution in Section~\ref{subsec:ml_model}.

\subsection{Key Idea} \label{subsec:key_idea}
The mmWave/THz communication systems suffer from severe path loss, which makes line-of-sight (LOS) communication between the basestation and user a preferable setting. This reliance on LOS communication draws an essential parallel with cameras, which also primarily capture visible or LOS objects. Further, these systems adopt large antenna arrays and use narrow directive beams to guarantee sufficient receive SNR. Selecting the optimal beams in these systems is typically associated with large beam training overhead, which makes it challenging to support highly-mobile users. Directing the beams can be viewed as focusing the signals in a particular direction in space. The beam vectors divide the scene (spatial dimensions) into multiple (possibly overlapping) sectors, where each sector is associated with a particular beam value. Therefore, given a pre-defined codebook, the beam prediction task can be transformed into a classification task, i.e., depending on the user's location in the visual scene, a beam index from the codebook can be assigned. All this motivated using visual data for beam prediction in mmWave drones. Further, the recent advancements in machine learning and computer vision have enabled several novel capabilities, such as object detection, image segmentation, and object tracking, to name a few. Such capabilities enable detecting the different objects of interest and extracting the relative position of the user in the visual scene. In this paper, instead of performing conventional beam training, we compute the optimal beam indices by utilizing the visual data captured by the camera installed at the mmWave drones.

\subsection{Machine Learning Model} \label{subsec:ml_model}
As presented in Section~\ref{sec:prob_form}, the objective is to learn the class-prediction function $f_\Theta(X)$ using the RGB images captured by the mmWave drones. The proposed vision-aided solution utilizes the state-of-the-art convolutional neural network (CNN) to predict the optimal beams. To realize the high reliability and low-latency requirements in mmWave communication, the CNN in the proposed solution must meet two essential criteria: (i) Accurate prediction and (ii) low inference latency. Unlike DNN networks such as VGG-Net, ResNet adopts residual blocks with skip connections that help achieve higher accuracy for image classification tasks with less number of parameters. Further, there was a significant decrease in the number of floating point operations per second (FLOPs) in ResNet compared to VGG-Net. Therefore, the proposed solution utilizes an ImageNet2012 pre-trained ResNet model \cite{resnet} for the beam prediction task. For a detailed comparison between accuracy and latency, the study considers two architectures: (i) A smaller 18-layer residual network (ResNet-18) and (ii) a larger 50-layer residual network (ResNet-50). The pre-trained ResNet models are modified to fit our beam prediction task, i.e., the final classification layer is replaced with a fully-connected layer with $Q$ output neurons. The intuition behind implementing transfer learning is that if a model is well-trained on a large and general enough dataset, it will efficiently generalize on other visual datasets. The ResNet models are further fine-tuned on the beam prediction dataset $\mathcal{D}$ consisting of image-beam pairs.

\subsection{Pruning Filters and Feature Maps}\label{sec:pruning}
DNNs with large capacities have many redundant parameters, i.e., filters and neurons. The number of matrix-multiplication in a single-forward pass directly correlates with the latency of the model. Along with achieving reliable performance, the proposed solution has to meet the mmWave communication systems' ultra-low-latency requirement. Therefore, to further reduce the computation cost of inference, we prune the redundant filters in the DNN as proposed in \cite{li2017pruning}. Compared to pruning weights, which introduce sparsity across the network, filter pruning is a structured way of pruning and reduces the FLOPS of the network. The goal is to prune the redundant filters from each layer of the well-trained model while minimizing the inference accuracy loss. The first step is to sample the filters based on the importance score, which measures each filter's impact on the final loss function. The relative importance score of a filter in each layer is calculated as the sum of its absolute weight, i.e., l1-norm of the filter. The filters with relatively smaller absolute weights tend to produce feature maps with weaker activation. Therefore, based on the importance score, $r$ percentage of the low-ranking filters are pruned. The resultant network is further fine-tuned to recover the lost accuracy due to pruning.

\section{ViWi-Drone Dataset}\label{sec:dataset}

The publicly available ViWi \cite{ViWi} framework is used to develop a new scenario specifically for the ViWi-Drone task. The dataset provides co-existing wireless and visual data. Each data sample contains an RGB image and the beam index. They are generated from an extensive simulation of a synthetic outdoor environment depicting a downtown street with multiple moving objects. Different from ViWi-BT, the user in the ViWi-Drone dataset is the drone flying at a height of $50$ meters. The drone is equipped with three cameras in different directions to capture the whole street at any given time instance and a half-wave dipole receiver antenna. Visual data from the drone cameras span the x-axis from $-100$ to $300$. The scenario also contains two basestations positioned on the opposite side of the street at either end of the main street, spaced $100$ meters apart. Each basestation is equipped with a half-wave dipole antenna array oriented along the z-axis with $128$ antennas along the x-axis. More information on the dataset can be found at \cite{ViWi}.

\begin{table}[!t]
	\caption{Design and Training Hyper-parameters}
	\centering
	\setlength{\tabcolsep}{5pt}
	\renewcommand{\arraystretch}{1.2}
	\begin{tabular}{|l|c|c|}
		\hline
		\multirow{6}{*}{Training}  & Batch Size                     & $128$                       \\ \cline{2-3} 
		& Weight decay                   & $1 \times 10 ^{-4}$ \\  \cline{2-3}                           
		& Learning Rate                  & $1 \times 10 ^{-4}$ \\  \cline{2-3} 
		& Learning Rate Decay            & epochs $10$ and $20$                       \\ \cline{2-3}   
		& Learning Rate Reduction Factor & $0.1$                       \\ \cline{2-3}                          
		& Number of Training Epochs               & $30$                         \\ \hline
	\end{tabular}
	\label{tab_nn_train_params}
\end{table}


\textbf{Drone Trajectories:} 
The dataset consists of $6,735$ data samples corresponding to $17$ different drone trajectories.The drone trajectories were generated using portions of the same car/human model trajectories that were created to animate the traffic used in the outdoor scene. Each trajectory only differs in the x and y-axis. The z-axis, i.e., the height of the flight, is kept fixed at 50 meters. The drones travel in a linear path either in the positive and negative x direction. The paths are located directly above the street that vary along the y axis from $-5.625$ to $1.875$. The $6,735$ data samples, i.e., image and beam pairs, are further divided into two subgroups: (i) dataset consisting of image and beam pairs for basestation 1 only, referred to as the BS1 scenario for further analysis; (ii) data samples corresponding to basestation 2 only, and is referred to as BS2 scenario. The combined dataset referred to as combined BS1 and BS2 scenario consists of samples from both basestation 1 and 2. For each of the three scenarios, the dataset is split into training and validation sets, using a $70\%-30\%$ data split.

\begin{figure}[t]
	\centering
	\subfigure[ResNet-18]{\centering \includegraphics[width=1.0\linewidth]{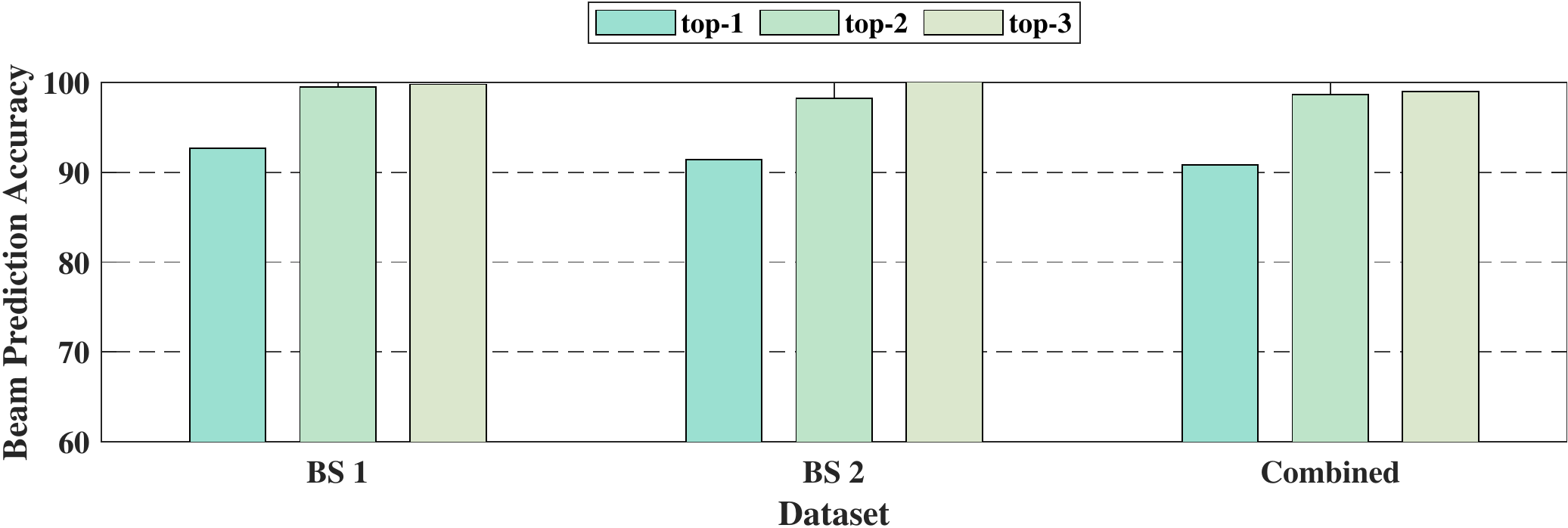}\label{fig:resnet18_acc}}
	\subfigure[ResNet-50]{\centering \includegraphics[width=1.0\linewidth]{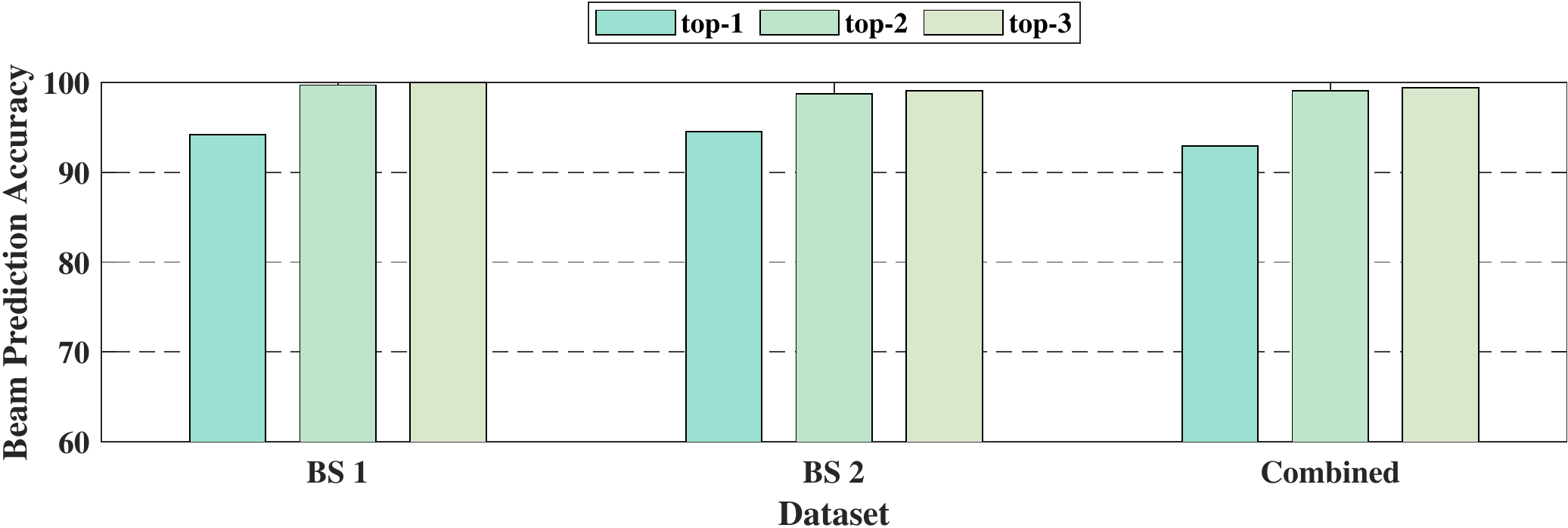}\label{fig:resnet50_acc}}
	\caption{This figure plots the top-k beam prediction accuracies (k $\in (1,2,3)$) for the proposed vision-aided solution. Fig. (a) and (b) present the top-k accuracies for ResNet-18 and ResNet-50 models, respectively. }
	\label{fig:topk_acc}
\end{figure}


\begin{figure*}[!t]
	\centering
	\includegraphics[width=0.85\linewidth]{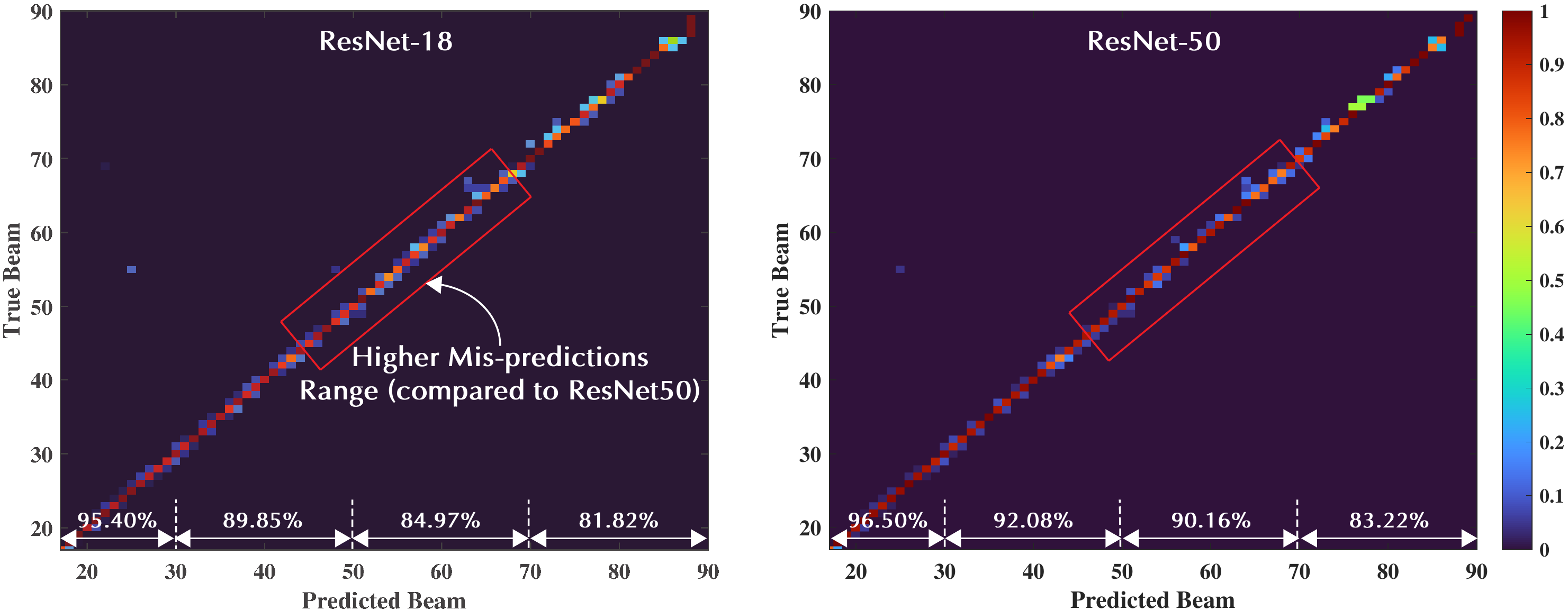}
	\caption{This figure plots the confusion matrices of the top-1 predicted beam indices (combined dataset) for both ResNet-18 and ResNet-50 models. It further shows the beam prediction accuracies for the different beam ranges.  }
	\label{fig:confusion_matrix}
\end{figure*}

\begin{figure}[!t]
	\centering
	\includegraphics[width=0.85\linewidth]{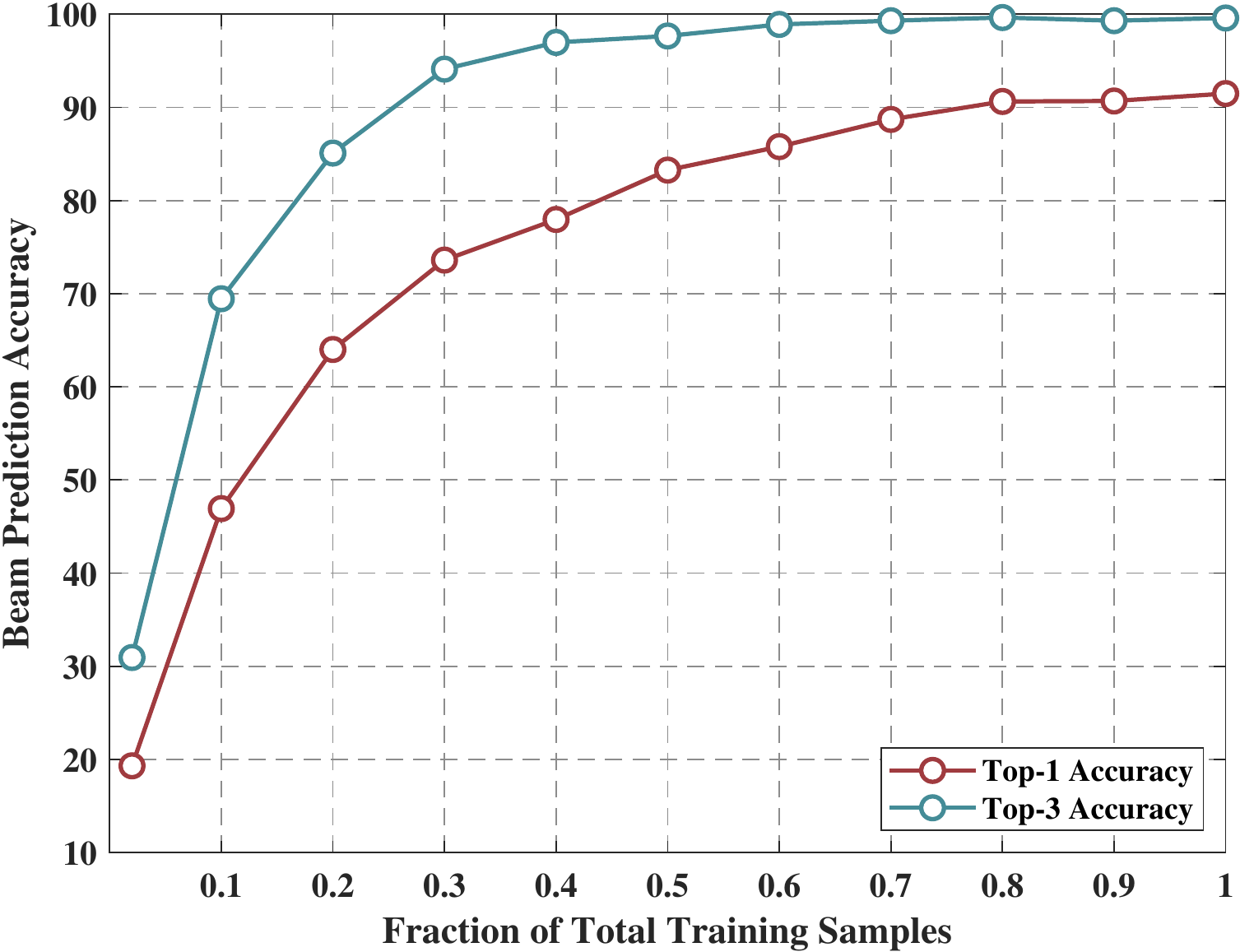}
	\caption{This figure shows the top-1 and top-3 beam prediction accuracy for the proposed vision-position model versus different dataset sizes. }
	\label{fig:percent_samples}
\end{figure}

\section{Experimental Setup}\label{sec:exp_set}
In this section, we present the details of our experimental setup. We first discuss the neural network training parameters followed by the adopted evaluation metric. 
 
\textbf{Network Training: } As described in Section~\ref{subsec:ml_model}, we validate our proposed method of image-based beam tracking on two different ImageNet2012 pre-trained ResNet architectures, i.e., ResNet-18, and ResNet-50. Both the models are further modified by removing the last fully-connected layer and replacing it with a fully-connected layer of size 128 neurons, i.e., $Q=128$. The cross-entropy loss with the Adam optimizer is used to train the models.The hyper-parameters used to fine-tune the model are presented in Table~\ref{tab_nn_train_params}. All the experiments were performed on a single NVIDIA RTX TITAN GPU using the PyTorch framework.

\textbf{Evaluation Metrics:}
In this section, we present the evaluation metric followed to evaluate the performance of our proposed network architecture. The primary method of evaluating model performance is using top-k accuracy. The top-k accuracy is defined as the percentage of the test samples where the ground-truth beam is within the top-$k$ predicted beams. The proposed solution needs to be highly accurate and still meet the low latency requirement. The mmWave communication demands near real-time inference capabilities, usually within tens of milliseconds for each sample. Therefore, along with the prediction accuracy we incorporate a comprehensive study of the post-training inference latency of our proposed models. 

\section{Performance Evaluation} \label{sec:perf_eval}
The proposed model is evaluated on the three scenarios, as mentioned in Section~\ref{sec:dataset}, i.e., (i) Basestation 1, (ii) Basestation 2, and (iii) Combined BS1 and BS2. The following two sections discuss the performance evaluation of the proposed ResNet-18 and ResNet-50 model in terms of beam prediction performance and inference latency.  

\subsection{Beam Prediction Performance}\label{sec:pred_acc}
In this section, we evaluate the performance of the proposed vision-aided solution from different perspectives such as the beam prediction accuracy, number of training samples, etc. 

\textbf{Can visual data help in predicting the optimal beam indices?} Fig.~\ref{fig:resnet18_acc} and Fig.~\ref{fig:resnet50_acc} plots the top-1, top-2, top-3 beam prediction accuracy for the three different scenarios (BS1, BS2, and Combined) with ResNet-18 and ResNet-50 models, respectively. It is observed that for all three scenarios, the proposed solution can achieve a high top-1 beam prediction accuracy of more than $90\%$. The performance improves further as we consider the top-2 and top-3 predicted beams, with beam prediction accuracy of close to $100\%$. As shown in Fig.~\ref{fig:img_samples}, we consider a realistic scenario with multiple objects in the environment, such as buses, trucks, cars, etc. The higher beam prediction accuracy highlights that additional sensory data, such as vision, can be utilized to predict the optimal beams with high fidelity. It is observed that the ResNet-50 model achieves better beam prediction performance for all three scenarios. This improvement in performance can be attributed to the higher number of parameters in the ResNet-50 model compared to that of the ResNet-18 model. It is important to note here that the number of parameters in ResNet-50 (23M) is almost twice that of ResNet-18 model (11M). However, the overall performance improvement is in the range of $1-2\%$, highlighting the trade-off between performance and computational cost. Further, we observe that both the models achieve high beam prediction accuracies for the more challenging combined dataset. Although, there is a slight drop in accuracy, both the networks achieve close to optimum beam prediction accuracy.

\textbf{Is the prediction accuracy uniform across different beams? } In Fig.~\ref{fig:confusion_matrix}, we plot the confusion matrices of the top-1 predicted beam indices (combined dataset) for both ResNet-18 and ResNet-50 models. This figure shows that both models can predict the optimal beam indices accurately. It also highlights that even in the case of mispredictions, the predicted beam is close to the ground-truth beam. Next, we compute the beam prediction performance for the different beam ranges. For instance, as shown in Fig.~\ref{fig:confusion_matrix}, the prediction accuracies of indices between $30$ and $50$ are $89.85\%$ and $92.08\%$ for ResNet-18 and ResNet-50, respectively. It is observed that for the lower ranges of the beam indices, the performance of both models is similar. However, for beam indices between $50$ and $70$, the performance of the ResNet-50 model surpasses that of ResNet-18 by almost $6\%$. This further shows that the ResNet-50 model with a higher number of parameters can predict the beams accurately for more challenging samples, such as when both the basestations are visible in the same image sample. The performance of beam indices between $70$ and $90$ is low primarily due to the lower number of samples in the training dataset for this range. 


\begin{table}[!t]
	\caption{Inference Latency Analysis of ResNet-18}
	\centering
	\setlength{\tabcolsep}{5pt}
	\renewcommand{\arraystretch}{1.5}
	\begin{tabular}{|c|c|c|c|c|}
		\hline
		\multicolumn{1}{|c|}{\multirow{2}{*}{\textbf{\begin{tabular}[c]{@{}c@{}}Pruning\\ Percentage\end{tabular}}}} & \multicolumn{1}{l|}{\multirow{2}{*}{\textbf{Parameters}}} & \multicolumn{1}{l|}{\multirow{2}{*}{\textbf{Accuracy}}} & \multicolumn{2}{c|}{\textbf{Latency/Image}}                                              \\ \cline{4-5} 
		\multicolumn{1}{|c|}{}                                                                                       & \multicolumn{1}{l|}{}                                     & \multicolumn{1}{l|}{}                                   & \multicolumn{1}{l|}{\textbf{Batch-Size 1}} & \multicolumn{1}{l|}{\textbf{Batch-Size 10}} \\ \hline
		Baseline                                                                                                     & 11.2 M                                                    & 0.9268                                                  & 6.28 ms                                    & 0.96 ms                                     \\ \hline
		59.82\%                                                                                                      & 4.5 M                                                     & 0.9363                                                  & 4.87 ms                                    & 0.71 ms                                     \\ \hline
		73.21\%                                                                                                      & 3.0 M                                                     & 0.9344                                                  & 4.83 ms                                    & 0.56 ms                                     \\ \hline
		83.03\%                                                                                                      & 1.9 M                                                     & 0.9287                                                  & 4.49 ms                                    & 0.49 ms                                     \\ \hline
		88.39\%                                                                                                      & 1.3 M                                                     & 0.9154                                                  & 4.43 ms                                    & 0.44 ms                                     \\ \hline
		95.08\%                                                                                                      & 0.55M                                                     & 0.9139                                                  & 4.36 ms                                    & 0.43 ms                                     \\ \hline
	\end{tabular}
	\label{tab:pruning}
\end{table}


\textbf{How many data samples are needed to predict the optimal beam indices?} Next, we draw some insights into the required dataset size. To do that, we present in \figref{fig:percent_samples} the top-1 and top-3 beam prediction accuracies versus the fraction of the total training dataset size. Note that the total size of the real-world dataset described in \sref{sec:dataset} is around $6,735$ data points divided into 70\% training and 30\% validation. \figref{fig:percent_samples} shows that the proposed vision-aided solution is capable of achieving more than $80\%$ top-1 accuracy with just $50\%$ of the total training samples, i.e., around $2,350$ samples out of $4,715$ training samples. These results highlight the efficiency and potential of the proposed vision-aided approach for the beam prediction task in mmWave drones.

\subsection{Inference Latency}\label{subsec:inf_latency}
The final adoption of any solution depends on achieving both high accuracy and very-low latency. Although the proposed DNN-based solution achieves high prediction accuracy on the ViWi-Drone dataset, as shown in Section~\ref{sec:pred_acc}, the solution still needs to satisfy the low-latency requirement of mmWave communication. The inference latency of a neural network model is dependent on the total number of matrix-multiplication operations required in a forward pass. To reduce the computations in the forward pass, we perform filter-based pruning as described in Section~\ref{sec:pruning}. In Table~\ref{tab:pruning}, we present the impact of pruning on inference latency. The analysis was performed on the proposed ResNet-18 architecture using an $8$GB NVIDIA RTX $2070$ GPU. It is observed that $\approx$ 95\% of the filters can be pruned without a significant drop in accuracy resulting in $\approx 30\%$ improvement in inference latency. This inherent trade-off between inference accuracy and achievable latency must be considered during network design.

\section{Conclusion}\label{sec:conc}

This paper develops a deep learning-based solution that utilizes visual data to predict the optimal beam indices in mmWave/THz drone communication systems. TO evaluate the proposed solution, we adopt the ViWi-Drone dataset consisting of diverse wireless and visual data. The proposed vision-aided solution achieves a top-$1$ and top-$3$ beam prediction accuracies of $\approx 91\%$ and $100\%$. We, further, show that by adopting state-of-the-art network pruning approaches, the proposed solution can achieve near real-time prediction latency of $4.36$ms without significant reduction in the prediction accuracy. The experimental results indicate the promising gains of leveraging additional sensory data such visual data (RGB images) to reduce the beam training overhead in mmWave/THz drones.


\end{document}